\documentclass[a4paper,11pt]{article}
\usepackage[dvips]{graphicx}
\usepackage{amssymb}
\usepackage{amsmath}

\newcommand{\plb}{Phys. Lett. B. }

\newcommand{\arx}{arXiv:}

\begin{document}

\title{Casimir Pistons with Curved Boundaries}
\author{V.K.Oikonomou\thanks{
voiko@physics.auth.gr}\\
Dept. of Theoretical Physics Aristotle University of Thessaloniki,\\
Thessaloniki 541 24 Greece\\
and\\
Technological Education Institute of Serres, 62124 Serres, Greece}
\maketitle

\begin{abstract}
In this paper we study the Casimir force for a piston
configuration in $R^3$ with one dimension being slightly curved
and the other two infinite. We work for two different cases with
this setup. In the first, the piston is ''free to move'' along a
transverse dimension to the curved one and in the other case the
piston ''moves'' along the curved one. We find that the Casimir
force has opposite signs in the two cases. We also use a
semi-analytic method to study the Casimir energy and force. In
addition we discuss some topics for the aforementioned piston
configuration in $R^3$ and for possible modifications from extra
dimensional manifolds.
\end{abstract}

\bigskip
\bigskip
\section*{Introduction}

More than 60 years have passed since H. Casimir's \cite{Casimir}
originating paper, stating that there exist's an attractive force
between two neutral parallel conducting plates. After 20 years,
the scientific society had appreciated this work and extended this
work in various areas, from solid state physics to quantum field
theory and even cosmology \cite{Elizalde,eli,Bordagreview}. The
Casimir effect is closely related with the existence of zero point
quantum oscillations of the electromagnetic field in the case of
parallel conducting plates. The boundaries polarize the vacuum and
that results to a force acting on the boundary
\cite{Bordagreview}. The Casimir force can be either repulsive or
attractive. That dependents on the nature of the background field
in the vacuum, the geometry of the boundary, the dimension and the
curvature of the spacetime. The regularization of the Casimir
energy is of particular important in order physical results become
clearer.

\noindent One very interesting configuration is the so called
Casimir piston. This configuration was originally treated
\cite{calvacanti} as a single rectangular box with three parallel
plates. The one in the middle is the piston. The dimensions of the
piston are $(L-a)\times b$ and $a\times b$, with the piston being
located in $a$. In \cite{calvacanti} the Casimir energy and
Casimir force for a scalar field was calculated. The boundary
conditions on the 'plates' where Dirichlet. There exists a large
literature on the subject
\cite{KirstenPistons,Teo,teo1,teo2,ElizaldePistons,Cheng,KirstenKaluzaKleinPiston,fulling,edery1,ar1,ar2,ar3}
calculating the Casimir force for various piston configurations
and for various boundary conditions of the scalar field. Also most
results where checked at finite temperature. The configurations
used where extended to include extra dimensional spaces which form
a product spacetime with the piston topology, that is
$M_{Piston}\times M^n$, with $M_{Piston}$ and $M^n$ the piston
spacetime topology and the extra dimensional spacetime topology.
In addition to the known Neumann and Dirichlet boundary
conditions, in reference \cite{ElizaldePistons} Robin boundary
conditions where considered for a Kaluza-Klein piston
configuration. In most cases the scalar field was taken massless
but there exists also literature for the massive case
\cite{TeoMassive}.

\noindent The calculational advantage the Casimir piston provides
is profound. Particularly, when one calculates the Casimir energy
between parallel plates confronts infinities that must be
regularized. The regularization of the Casimir energy in the
parallel plate geometry can be done if we calculate it as a sum
over discrete modes (due to boundary conditions on the plates)
minus the continuum integral (with no plates posing boundary
conditions)\cite{Bordagreview,Elizalde,eli}. The discrete sum
consists of three parts, a volume divergent term (which be
cancelled by the continuum integral), a surface divergent term and
a finite part. This can be easily seen if the calculations of the
Casimir energy are done with the introduction of a UV cutoff,
$\lambda$. Before the piston setup was firstly used, the surface
divergent term was thrown out. Actually this was proven to be
completely wrong because such surface divergent term cannot be
removed by renormalization of the physical parameters of the
theory \cite{jaffenew}. The zeta regularization technique
renormalizes this term to zero. Thus the cutoff technique and the
zeta regularization technique agree perfectly. However there is no
reason to justify the loss of the surface term within the cutoff
regularization technique. The Casimir piston solves this problem
in a very nice way, because the surface terms of the two piston
chambers cancel and thus the Casimir force can be consistently
calculated
\cite{KirstenPistons,Teo,teo1,teo2,ElizaldePistons,Cheng,KirstenKaluzaKleinPiston,fulling,edery1,ar1,ar2,ar3}.
We shall use the zeta-function regularization technique and also
treat the problem numerically. Also in the last section we shall
briefly give the expression of the Casimir energy using a cutoff
regularization technique and demonstrate how the surface terms
cancel.

\begin{figure}[h]
\begin{center}
\includegraphics[scale=.6]{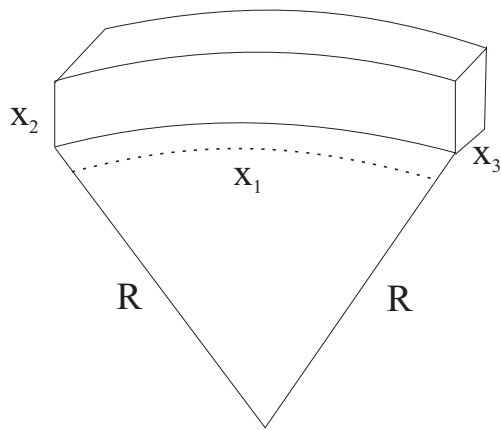}
\end{center}
\caption{Curved Piston Configuration with Infinite Radius}
\label{General Piston}
\end{figure}

\begin{figure}[h]
\begin{center}
\includegraphics[scale=.8]{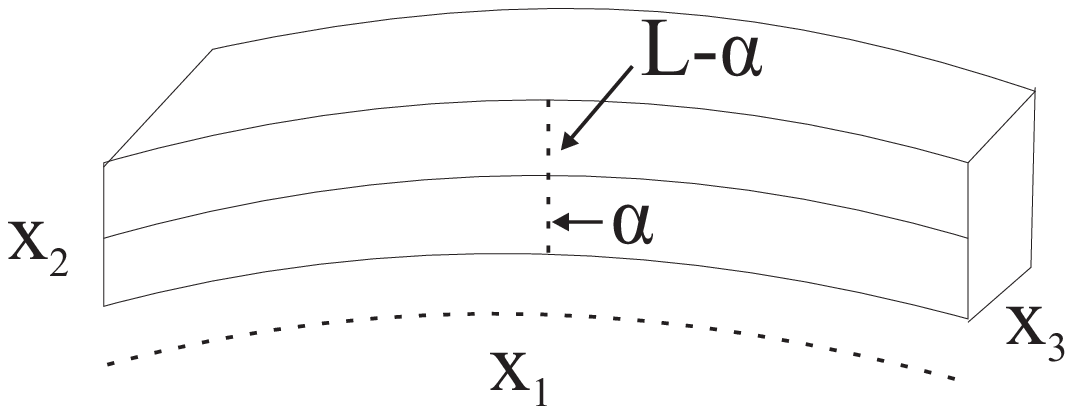}
\end{center}
\caption{\textbf{Piston 1}, The moving piston along the non-curved
dimension $x_2$} \label{MyPiston}
\end{figure}

\begin{figure}[h]
\begin{center}
\includegraphics[scale=.9]{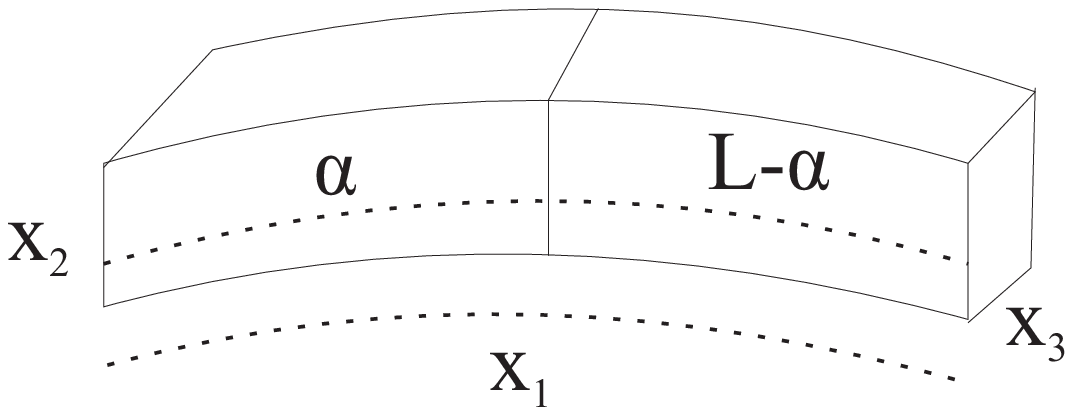}
\end{center}
\caption{\textbf{Piston 2}, The moving piston along the curved
dimension $x_1$} \label{MyPiston2}
\end{figure}

\noindent Motivated by a recent paper
\cite{KirstenKaluzaKleinPiston}, we shall consider a piston
configuration in $R^3$ space without the extra dimensions. In this
topology we consider a piston geometry living in three space. We
shall study two configurations which can be seen in Figures
\ref{MyPiston} and \ref{MyPiston2}. In the first case Fig.
\ref{MyPiston}, the $x_1$ and $x_3$ dimensions are sent to
infinity and we have ''parallel plates'' having distance $\L$
between them. The piston consists of two such three dimensional
chambers with the piston plate having distance $\alpha$ and
$L-\alpha$ from the boundary plates. In addition to this we shall
assume that in one of the two infinite dimensions, there is
curvature. As can be seen the infinite dimension $x_1$ seems to be
a part of a circle. We shall work in the case where the radius of
the circle goes to infinity, that is when $R\rightarrow \infty$.
We shall give the solution for the Laplace equation for a scalar
field (with Dirichlet boundary conditions at all boundaries, see
below) corresponding to this configuration and find the
eigenfrequencies. Next we calculate the Casimir energy and the
Casimir force for each chamber and for the whole configuration
using standard techniques \cite{Bordagreview,Elizalde,eli}.
Additionally we study a similar to the above situation, described
by Figure \ref{MyPiston2}. Particularly in this case the piston is
free to move along the curved dimension and $x_2$ and $x_3$ are
infinite dimensions. These cases shall be presented in section 1.
In section 2 we discuss the results we found in section 1. In
section 3 we present a semi-analytic calculation of the Casimir
energy and Casimir force. Finally the conclusions with a
discussion follow.

\bigskip

\noindent Before closing this section it worths discussing what is
the motivation to study scalar field Casimir energy (and closely
connected to this, the Casimir piston configurations) and
particularly at four spacetime dimensions, that is 3+1, 3 space
and one time. The studies of Casimir effect in higher dimensions
through the effective action calculation
\cite{periodiko,periodiko1,periodiko2} are very interesting
especially when vacuum stability issues are addressed. Pistons
including extra dimensions serve in defining the impact of extra
dimensions in an observable way. So why studying pistons in 3+1
dimensions? The first reason comes from studies of Bose-Einstein
condensates \cite{edery2,ar4,ar5}. Particularly when calculating
the Casimir energy and pressure in a zero temperature homogeneous
weakly interacting dilute Bose-Einstein condensate in a parallel
plate geometry including Bogoliubov corrections, the leading order
term is identified with the Casimir energy of a massless scalar
field \cite{edery2,ar4,ar5}. This fact begins a new era in
experiments because the scalar field Casimir force can be measured
in Bose-Einstein condensates. This will be the first thing to
measure actually, because the leading order term is the scalar
Casimir force. In connection with these, the Bose-Einstein
condensates have very interesting prospects. Particularly,
Bose-Einstein condensates are known to provide an effective metric
and for mimicking kinematic aspects of general relativity, thus
probing kinematic aspects of general relativity
\cite{edery2,ar4,ar5}.

\noindent Another motivation for studying scalar Casimir energies
in 3+1 dimensions is the connection of the scalar Casimir energy
with the electromagnetic field Casimir energy in 3+1 dimensions
\cite{ambjorn} and also the 4-dimensional perfect conductor
Casimir energy of the electromagnetic field is identical with the
3-dimensional scalar field Casimir energy by dimensional reduction
\cite{edery3}. Let us discuss on these two in detail. The Casimir
energy of an electromagnetic field in the radiation gauge is
connected with the Dirichlet massless scalar field Casimir energy
with the general relation (see for example Ambjorn and Wolfram
\cite{ambjorn}),
\begin{equation}\label{revision1}
E_{A}(a_1,a_2,..,a_p;p;d)=(d-1)E_{\phi_D}(a_1,a_2,..a_p;p;d)+\sum_{i=1}^pE_{\phi_D}(a_1,a_2,..a_{i-1},a_{i+1},...,a_p;p-1;d-1)
\end{equation}
In the above equation, $p$ and $d$ are the dimensions of a a
hypercuboidal region with p sides of finite length $a_1,a_2,..a_p$
and $d-p$ sides with length $L\gg a_i$, ($i=1,2,..p$).

\noindent Of course the electromagnetic field Casimir force
calculations are most valuable in the case of nano devices, where
attractive Casimir forces may cause the collapse of the device.
This is known as stiction \cite{edery2,ar4,ar5,Teo,teo1,teo2}.

\noindent One strong motivation to study scalar and
electromagnetic Casimir energies in various topologies and
geometries comes from the fact that the scalar and electromagnetic
Casimir energies are maybe connected quantitatively within the
piston setup. In a recent study, A. Edery et.al \cite{edery3}
studied exactly this issue. Let us discuss some of their results
relevant to us. The authors found that the original Cavalcanti's
2+1 dimensional scalar Dirichlet Casimir piston can be obtained by
dimensionally reducing a 3+1 dimensional electromagnetic piston
system obeying perfect conductor conditions. This novel
observation serves to open new directions in studying various
piston configurations in 3+1 dimensions. Having in mind that the
electromagnetic Casimir energy is closely related to the Dirichlet
scalar Casimir energy \cite{ambjorn}, this can be proven really
valuable. We must mention that as proved by the authors of
\cite{edery3}, this happens only for 3+1 dimensional
electromagnetic piston systems. This makes 3+1 dimensional piston
studies really interesting both theoretically and maybe at some
point experimentally, for example in nano devices where the
boundaries form the various topologies. We have to mention that
the dimensional reduction as worked out by the authors of
\cite{edery3} does not refer to the reduction of a toroidal
dimension but to the reduction of an interval (for example
$(0,L)$) with boundary conditions that respect the symmetries of
the original action (perfect electric conductor). The reduction is
obtained by setting the interval length go to zero, that is
$L\rightarrow 0$. They studied the general case with $d$
dimensions. In general the dimensional reduction (as in the
compact dimensions case) leads to a massive tower of massive
Kaluza-Klein states with mass inverse proportional to $L^2$. When
$L\rightarrow 0$ these states decouple from the theory only when
the original electromagnetic system leaves in 3+1 dimensions. The
initial search was to see whether a $d$ dimensional
electromagnetic field reduces to the $d-1$ electromagnetic field
after dimensional reduction. They found that for three dimensions
the electromagnetic field has one degree of freedom and also the
scalar field has one degree of freedom. Thus as long the boundary
conditions match, these two can be considered equivalent.
Particularly the Casimir force is identical in the two cases (see
\cite{edery3} for many details and the proof). This is a valuable
result both theoretically and experimentally. Experimentally
because this reduction scenario could be investigated in high
precision Casimir experiments with real metals in 3 space
dimensions (3+1 spacetime)

\noindent An additional theoretical prospect of the dimensional
reduction setup of reference \cite{edery3} is the similarity of
the whole configuration with orbifold boundary conditions. We
shall not discuss on this now but we hope to comment soon.

\noindent In addition, another result found in paper \cite{edery3}
is that the reduction of a $d$ dimensional Dirichlet scalar
Casimir piston does not lead to the $d-1$ dimensional scalar
Casimir piston. Actually under the reduction, the Casimir energy
goes to zero. This is intriguing showing us that the close
relation of electromagnetic 3+1 pistons with 2+1 scalar pistons is
really valuable. Regarding the $d$-dimensional scalar and it's
relation with the $d-1$ scalar theory there is a well known
relation between the two through finite temperature field theory
\cite{largen}. Specifically a theory at finite temperature offers
the possibility to connect a $d-1$ dimensional scalar theory with
the $d$ dimensional theory at finite temperature. However we must
be really cautious because the argument that a $d$ dimensional
field theory correspond to the same theory in $d-1$ dimensions has
been proven true \cite{largen} only for the $\phi^4$ theory
(always within the limits of perturbation theory). Also this also
holds true for supersymmetric theories. On the contrary this does
not hold for $QCD$ and Yang-Mills theories. Actually $QCD_3$
resembles more $QCD_4$ and not $QCD_4$ at finite temperature.
Following these lines it would be interesting to see whether this
argument holds for the scalar Casimir pistons with various
boundary conditions, that is to check whether the 3+1 dimensional
scalar Casimir piston force at finite temperature is equal to the
2+1 Casimir piston, in the infinite temperature limit. Also there
exists an argument stating that a $d$ dimensional theory at finite
temperature resembles more the same theory with one dimension
compactified to a circle and in the limit $R\rightarrow 0$, where
$R$ the magnitude of the compact dimension. It would be
interesting to examine this within the Edery's \cite{edery3}
dimensional reduction scheme.

\noindent Finally a motivation to study the piston configuration
we use in this paper is the asymmetry that one of the two piston
we use has. Particularly the figure \ref{MyPiston} configuration.
There exists an argument in the literature stating that the
Casimir force between bodies related by reflection is always
attractive, independent of the exact form of the bodies or
dielectric properties. As we will see in the following this
argument applies both to the two cases we use, thus proving that a
small deviation from the reflection symmetry does not modify the
standard results of rectangular pistons.

\section{The Piston Setup}

The configuration of Figure \ref{MyPiston} can be described as
follows: The curvature of the infinite length curved dimension is
$\frac{1}{R}$ and the width between plates is $\alpha$ and
$L-\alpha$ for the two chambers. We shall treat only the $\alpha$
chamber first. The generalization to the other case is
straightforward. In order to describe this slightly curved piston
setup more efficiently, we choose the coordinates $s,t$ to
describe the plane $x_1-x_2$ and $x_3$ remains the same, as it
appears in Figure \ref{MyPiston}. The piston is based on the
papers \cite{jaffe,jaffe1}. We borrowed the analysis and the
ansatz solutions that the authors used for bent waveguides.
Following \cite{jaffe,jaffe1}, the local element in the plane
$s-t$ is $\mathrm{d}A=h(s,t)\mathrm{d}s\mathrm{d}t$, with $s$ the
length along the infinite slightly curved dimension and $t$ the
transverse dimension, with $h(s,t)=1-\frac{1}{R}t$. The Laplacian
in terms of $s,t,x_3$ looks like (acting on $\psi$, with $\psi$ a
scalar field),
\begin{equation}\label{laplacian}
\nabla ^2 \psi =\frac{1}{h}\frac{\partial}{\partial
t}(h\frac{\partial \psi}{\partial
t})+\frac{1}{h}\frac{\partial}{\partial s}(h\frac{\partial
\psi}{\partial s})+\frac{\partial ^2 \psi}{\partial x_3 ^2}.
\end{equation}
If in the above relation we expand $h(s,t)$ for $R\rightarrow
\infty$, then the non-vanishing terms in first order approximation
yield the usual Laplacian in Cartesian coordinates. An ansatz
solution of $\nabla ^2 \psi =0$ is:
\begin{equation}\label{solution}
\psi (s,t,x_3)=\frac{u(s)}{\sqrt{h(s,t)}}\sin{(\frac{n\pi
t}{\alpha})}e^{ik_3{\,}x_3},
\end{equation}
subject to the Dirichlet boundary conditions, $\psi (s,t,x_3)=\psi
(s,t+\alpha ,x_3)=0$ for the transverse coordinate $t$. Relation
(\ref{solution}) for infinite $R$ yields (keeping first order non
vanishing terms):
\begin{equation}\label{solution1}
\psi (s,t,x_3)=u(s)\sin{\frac{n\pi t}{\alpha}}e^{ik_3{\,}x_3},
\end{equation}
with $u(s)$ satisfying:
\begin{equation}\label{onedimensional}
\frac{\mathrm{d}^2u}{\mathrm{ds}^2}+(k^2-\frac{\pi^2}{\alpha
^2}+\frac{1}{R^2})u(s)=0.
\end{equation}
Then the eigenfrequencies of the above configuration is,
\begin{equation}\label{eigenfrequencies}
\omega^2=k^2_1+k_2^2+(\frac{n\pi}{\alpha})^2-\frac{\pi^2}{\alpha
^2}+\frac{1}{R^2}.
\end{equation}
Notice that in the limit $R\rightarrow \infty$, the term
$-\frac{\pi^2}{\alpha ^2}+\frac{1}{R^2}$ is negative. In the next
section we shall comment on this.

\begin{figure}[h]
\begin{center}
\includegraphics[scale=.6]{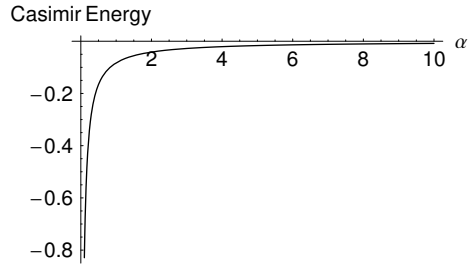}
\end{center}
\caption{The Casimir Energy for the chamber $\alpha$, for
$R=10^{100}$, $L=10^{11}$ (\textbf{Piston 1 Case})}
\label{CasimirEnergy1}
\end{figure}

\begin{figure}[h]
\begin{center}
\includegraphics[scale=.6]{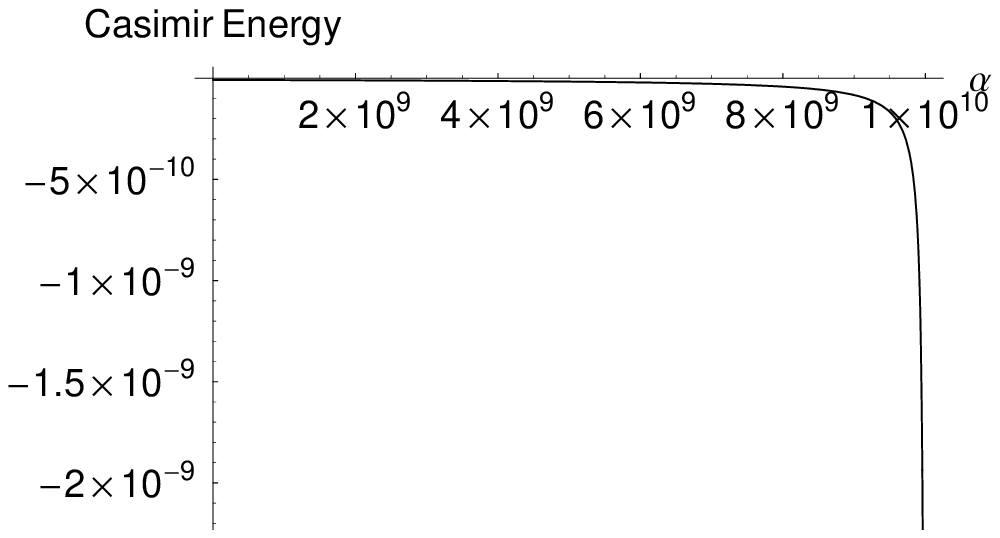}
\end{center}
\caption{The Casimir Energy for the chamber $L-\alpha$, for
$R=10^{100}$, $L=10^{11}$ (\textbf{Piston 1 Case})}
\label{CasimirEnergy2}
\end{figure}

\begin{figure}[h]
\begin{center}
\includegraphics[scale=.6]{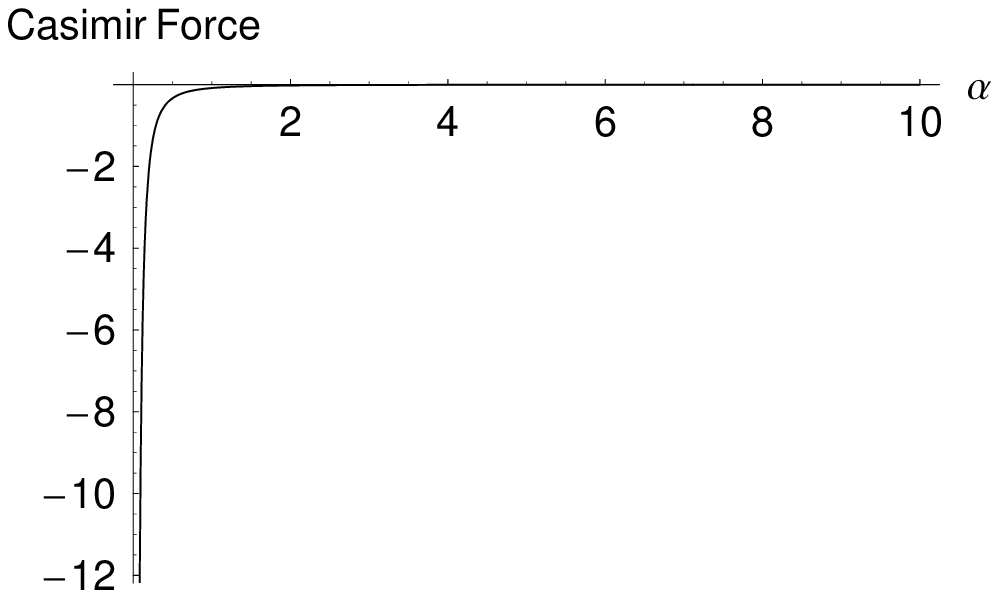}
\end{center}
\caption{The Casimir Force for the chamber $\alpha$, for
$R=10^{100}$, $L=10^{11}$ (\textbf{Piston 1 Case})}
\label{CasimirForce1}
\end{figure}

\begin{figure}[h]
\begin{center}
\includegraphics[scale=.6]{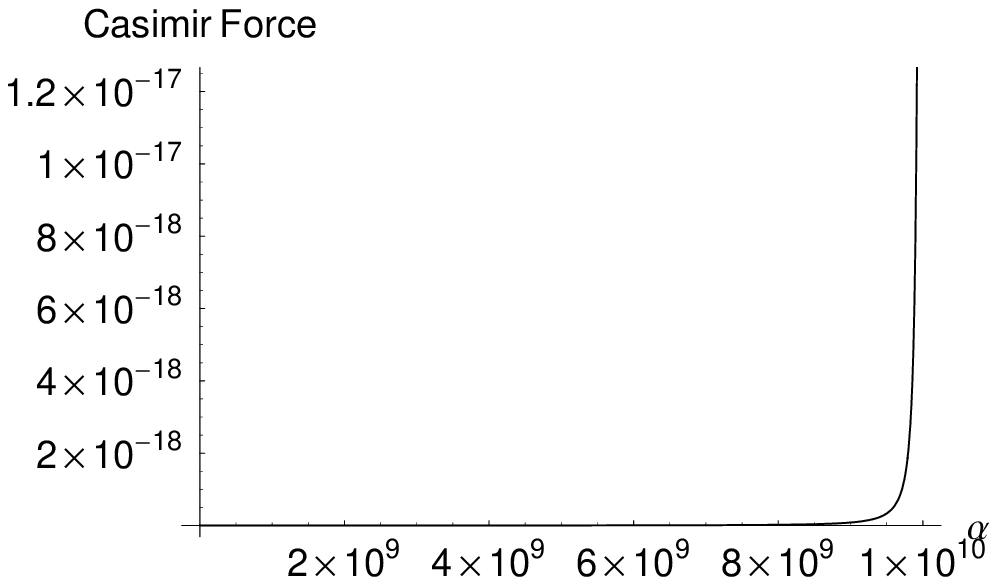}
\end{center}
\caption{The Casimir Force for the chamber $L-\alpha$, for
$R=10^{100}$, $L=10^{11}$ (\textbf{Piston 1 Case})}
\label{CasimirForce2}
\end{figure}

\begin{figure}[h]
\begin{center}
\includegraphics[scale=.6]{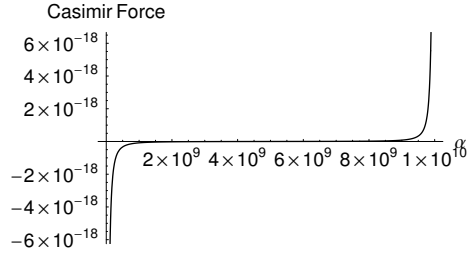}
\end{center}
\caption{The total Piston Casimir Force, for $R=10^{100}$,
$L=10^{11}$ (\textbf{Piston 1 Case})} \label{CasimirForceTotal}
\end{figure}

\noindent Now the Casimir energy per unit area for the chamber
with transverse length $\alpha$ is \cite{Bordagreview} (bearing in
mind that $-\frac{\pi^2}{\alpha ^2}+\frac{1}{R^2}$ is much less
than $1$):
\begin{align}\label{alfacasimir}
&E_{ren}(\alpha)= \\ \notag &\int_{-\infty}^{\infty}
\frac{\mathrm{d}k_1\mathrm{d}k_2}{4\pi^2}
\Big{(}\sum_{n=1}^{\infty}\sqrt{k^2_1+k_2^2+(\frac{n\pi}{\alpha})^2-\frac{\pi^2}{\alpha
^2}+\frac{1}{R^2}}
\\ \notag &-\frac{a}{\pi}\int_{\sqrt{-\frac{1}{R^2}+\frac{\pi^2}{\alpha
^2}}}^{\infty}\sqrt{k^2_1+k_2^2+(\frac{n\pi}{\alpha})^2-\frac{\pi^2}{\alpha
^2}+\frac{1}{R^2}}\Big{)},
\end{align}
where the left term is the finite $\alpha$ part and the right term
the infinite $\alpha$ part. The substraction of the continuum part
from the finite part corrections is a very well known
renormalization technique for the Casimir energy
\cite{Bordagreview}. Now, performing the $k_1$, $k_2$ integration,
one obtains:
\begin{align}\label{alfacasimir1}
&E_{ren}(\alpha)= \\ \notag & \frac{3}{2\pi}
\Big{(}\sum_{n=1}^{\infty}\big{[}(\frac{n\pi}{\alpha})^2-\frac{\pi^2}{\alpha
^2}+\frac{1}{R^2}\big{]}^{3/2}-\frac{a}{\pi}\int_{\sqrt{-\frac{1}{R^2}+\frac{\pi^2}{\alpha
^2}}}^{\infty}\big{[}(\frac{n\pi}{\alpha})^2-\frac{\pi^2}{\alpha
^2}+\frac{1}{R^2}\big{]}^{3/2}\Big{)}.
\end{align}
The $k_1$, $k_2$ integration is done using,
\begin{equation}\label{zetaintegrationk1}
\int \mathrm{d}k_1\mathrm{d}k_2\sum
_{n=1}^{\infty}\big{[}k_1^2+k_2^2+\big{(}\frac{n\pi}{a}\big{)}^2-m^2\big{]}^{-s},
\end{equation}
which after integrating over $k_1$ and $k_2$ we obtain,
\begin{equation}\label{k1k2}
\Gamma (s-1){\,}\pi\sum _{n=1}^{\infty}
\big{[}\big{(}\frac{n\pi}{a}\big{)}^2-m^2\big{]}^{1-s}.
\end{equation}
Analytically continuing the above to $s=-\frac{1}{2}$ we obtain
relation (\ref{alfacasimir1}). Upon using the Abel-Plana formula
\cite{Bordagreview,Saharian},
\begin{equation}\label{AbelPlana}
\sum_{n=0}^{\infty}f(n)-\int_{0}^{\infty}f(n)\mathrm{d}n=\frac{1}{2}f(0)+i\int_{0}^{\infty}\mathrm{d}t\frac{f(it)-f(-it)}{e^{2\pi
t}-1},
\end{equation}
the Casimir energy (\ref{alfacasimir1}) becomes,
\begin{align}\label{alfacasimirplana}
&E_{ren}(\alpha)=\frac{3}{2\alpha}\Big{(}\frac{1}{2}(1-b^2)^{3/2}
\\ \notag &-\int_{0}^{\infty}\mathrm{d}t\frac{1}{e^{2\pi
t}-1}\Big{[}2t\sqrt{\frac{\sqrt{(1-b^2-t^2)^2+4t^2}+1-b^2-t^2}{2}}
\\ \notag &+(1-b^2-t^2)\sqrt{\frac{\sqrt{(1-b^2-t^2)^2+4t^2}-1+b^2+t^2}{2}}\Big{]}
\\ \notag &+\frac{1}{8}\big{(}(5b^2-2)\sqrt{-b^2+1}+3b^4(\ln b-\ln (1+\sqrt{1-b^2})) \big{)}
\Big{)},
\end{align}
with
\begin{equation}\label{beta}
b=\frac{\alpha\sqrt{-\frac{1}{R^2}+\frac{\pi^2}{\alpha ^2}}}{\pi}.
\end{equation}
Following the same steps us above, with $a\rightarrow L-\alpha$,
we obtain the renormalized Casimir energy for the chamber
$L-\alpha$:
\begin{align}\label{alfacasimirplanala}
&E_{ren}(L-\alpha)=\frac{3}{2(L-\alpha)}\Big{(}\frac{1}{2}(1-\lambda
^2)^{3/2}
\\ \notag &-\int_{0}^{\infty}\mathrm{d}t\frac{1}{e^{2\pi
t}-1}\Big{[}2t\sqrt{\frac{\sqrt{(1-\lambda
^2-t^2)^2+4t^2}+1-\lambda ^2-t^2}{2}}
\\ \notag &+(1-\lambda ^2-t^2)\sqrt{\frac{\sqrt{(1-\lambda ^2-t^2)^2+4t^2}-1+\lambda ^2+t^2}{2}}\Big{]}
\\ \notag &+\frac{1}{8}\big{(}(5\lambda ^2-2)\sqrt{-\lambda ^2+1}+3\lambda^4(\ln \lambda-\ln (1+\sqrt{1-\lambda ^2}{\,})) \big{)}
\Big{)},
\end{align}
and for this case
\begin{equation}\label{lampda}
\lambda=\frac{(L-\alpha)\sqrt{-\frac{1}{R^2}+\frac{\pi^2}{(L-\alpha)
^2}}}{\pi}.
\end{equation}
Now the total Casimir energy for the piston is obtained by adding
relations (\ref{alfacasimirplana}) and (\ref{alfacasimirplanala}),
namely,
\begin{equation}\label{totalcasimirenergy}
E_{Piston}=E_{ren}(L-\alpha)+E_{ren}(\alpha).
\end{equation}
In Figures \ref{CasimirEnergy1}, \ref{CasimirEnergy2} and we plot
the Casimir energy for the chamber $\alpha$, $L-\alpha$
respectively, for the numerical values $R=10^{100}$, $L=10^{11}$.
Also in Figures \ref{CasimirForce1}, \ref{CasimirForce2} and
\ref{CasimirForceTotal} we plot the Casimir force for the chambers
$\alpha$, $L-\alpha$ and the total Casimir force, respectively.

\subsection{Another Piston Configuration}

\begin{figure}[h]
\begin{center}
\includegraphics[scale=.6]{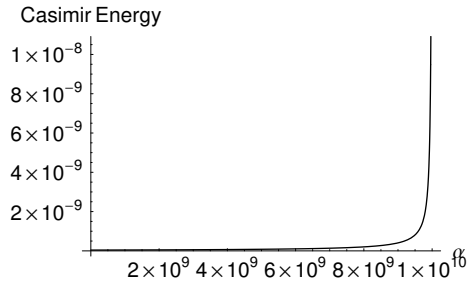}
\end{center}
\caption{The Casimir Energy for the chamber $\alpha$, for
$R=10^{100}$, $L=10^{11}$ (\textbf{Piston 2 Case})}
\label{Casimir1}
\end{figure}

\begin{figure}[h]
\begin{center}
\includegraphics[scale=.6]{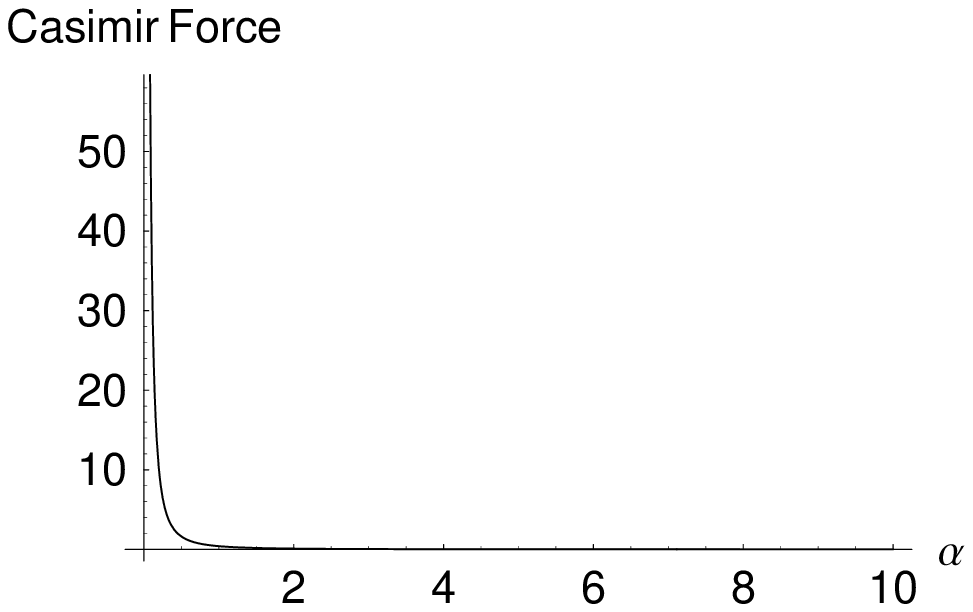}
\end{center}
\caption{The Casimir Force for the chamber $\alpha$, for
$R=10^{100}$, $L=10^{11}$ (\textbf{Piston 2 Case})}
\label{CasimirF1}
\end{figure}

\begin{figure}[h]
\begin{center}
\includegraphics[scale=.6]{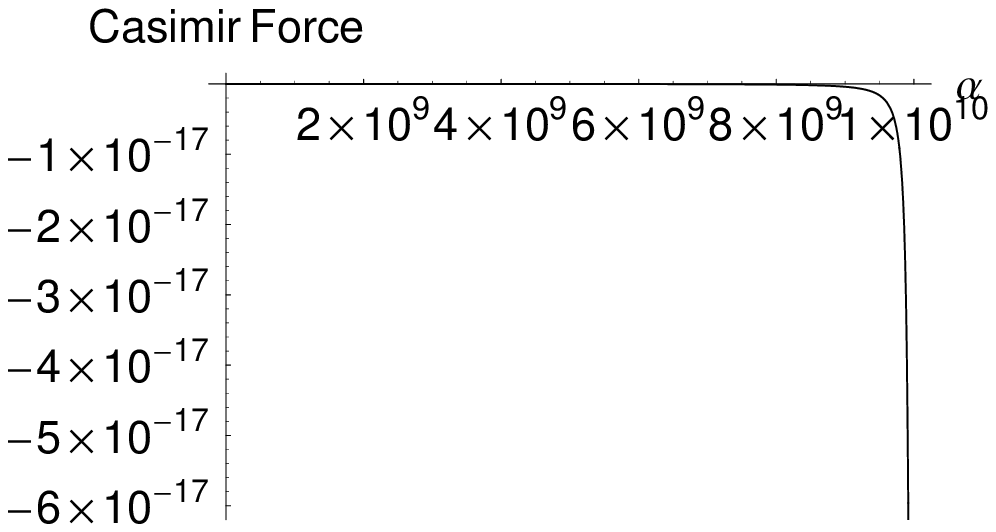}
\end{center}
\caption{The Casimir Force for the chamber $L-\alpha$, for
$R=10^{100}$, $L=10^{11}$ (\textbf{Piston 2 Case})}
\label{CasimirF2}
\end{figure}

Let us now study the case for the piston configuration appearing
in Figure \ref{MyPiston2}. Following the previous steps, the
eigenvalue spectrum of the Laplacian is:
\begin{equation}\label{pistoneigen2}
\omega ^2=k^2_1+k_2^2+(\frac{n\pi}{\alpha})^2-\frac{1}{R^2}.
\end{equation}
Now the Casimir energy for the $\alpha$ piston is (after
integrating on the infinite dimensions),
\begin{align}\label{alfacasimir12}
&E_{ren}(\alpha)= \\ \notag & \frac{3}{2\pi}
\Big{(}\sum_{n=1}^{\infty}\big{[}(\frac{n\pi}{\alpha})^2-\frac{1}{R^2}\big{]}^{3/2}-\frac{a}{\pi}\int_{\frac{1}{R}}^{\infty}\big{[}(\frac{n\pi}{\alpha})^2-\frac{1}{R^2}\big{]}^{3/2}\Big{)}.
\end{align}
Upon using again the Abel-Plana formula
\cite{Bordagreview,Saharian}, the Casimir energy
(\ref{alfacasimir12}) becomes,
\begin{align}\label{alfacasimirplana1}
&E_{ren}(\alpha)=\frac{3}{2\alpha}\Big{(}\frac{1}{2}(1-b^2)^{3/2}
\\ \notag &-\int_{0}^{\infty}\mathrm{d}t\frac{1}{e^{2\pi
t}-1}\Big{[}2t\sqrt{\frac{\sqrt{(1-b^2-t^2)^2+4t^2}+1-b^2-t^2}{2}}
\\ \notag &+(1-b^2-t^2)\sqrt{\frac{\sqrt{(1-b^2-t^2)^2+4t^2}-1+b^2+t^2}{2}}\Big{]}
\\ \notag &+\frac{1}{8}\big{(}(5b^2-2)\sqrt{-b^2+1}+3b^4(\ln b-\ln (1+\sqrt{1-b^2})) \big{)}
\Big{)},
\end{align}
with
\begin{equation}\label{beta2}
b=\frac{\alpha}{R\pi}.
\end{equation}
Following the same steps us above, with $a\rightarrow L-\alpha$,
we obtain the renormalized Casimir energy for the chamber
$L-\alpha$:
\begin{align}\label{alfacasimirplanala2}
&E_{ren}(L-\alpha)=\frac{3}{2(L-\alpha)}\Big{(}\frac{1}{2}(1-\lambda
^2)^{3/2}
\\ \notag &-\int_{0}^{\infty}\mathrm{d}t\frac{1}{e^{2\pi
t}-1}\Big{[}2t\sqrt{\frac{\sqrt{(1-\lambda
^2-t^2)^2+4t^2}+1-\lambda ^2-t^2}{2}}
\\ \notag &+(1-\lambda ^2-t^2)\sqrt{\frac{\sqrt{(1-\lambda ^2-t^2)^2+4t^2}-1+\lambda ^2+t^2}{2}}\Big{]}
\\ \notag &+\frac{1}{8}\big{(}(5\lambda ^2-2)\sqrt{-\lambda ^2+1}+3\lambda^4(\ln \lambda-\ln (1+\sqrt{1-\lambda ^2})) \big{)}
\Big{)},
\end{align}
and for this case
\begin{equation}\label{lampda1}
\lambda=\frac{(L-\alpha)}{R\pi}.
\end{equation}
Now the total Casimir energy for the piston is obtained by adding
relations (\ref{alfacasimirplana1}) and
(\ref{alfacasimirplanala2}), namely,
\begin{equation}\label{totalcasimirenergy1}
E_{Piston}=E_{ren}(L-\alpha)+E_{ren}(\alpha).
\end{equation}
In Figures \ref{Casimir1} we plot the Casimir energy for the
chamber $\alpha$ and for the numerical values $R=10^{100}$,
$L=10^{11}$. Also in Figures \ref{CasimirF1}, \ref{CasimirF2} and
\ref{CasimirFT1} we plot the Casimir force for the chamber
$\alpha$ the chamber $L-\alpha$ and the total Casimir force.
\begin{figure}[h]
\begin{center}
\includegraphics[scale=.6]{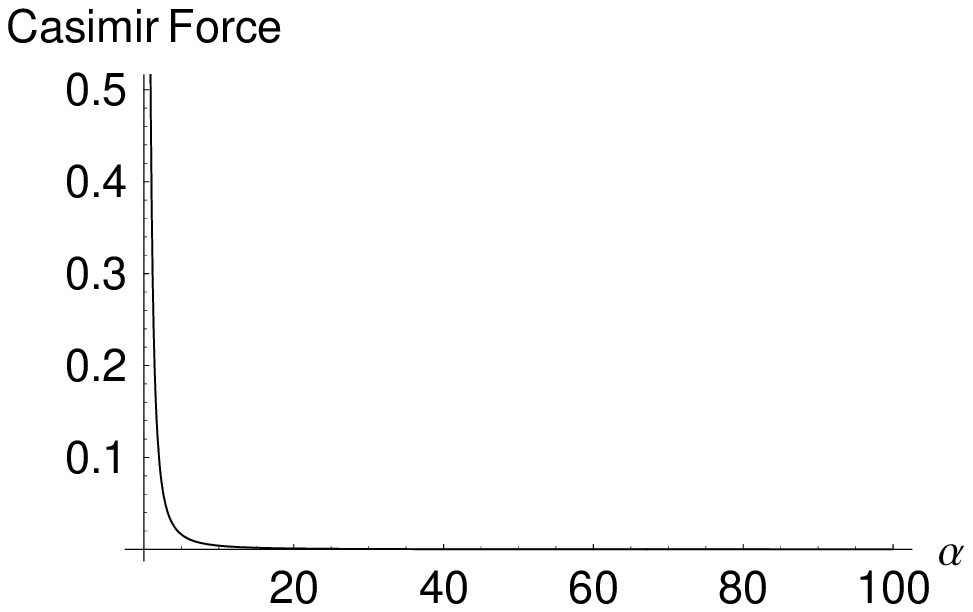}
\end{center}
\caption{The total Piston Casimir Force, for $R=10^{100}$,
$L=10^{11}$ (\textbf{Piston 2 Case})} \label{CasimirFT1}
\end{figure}
\begin{figure}[h]
\begin{center}
\includegraphics[scale=.6]{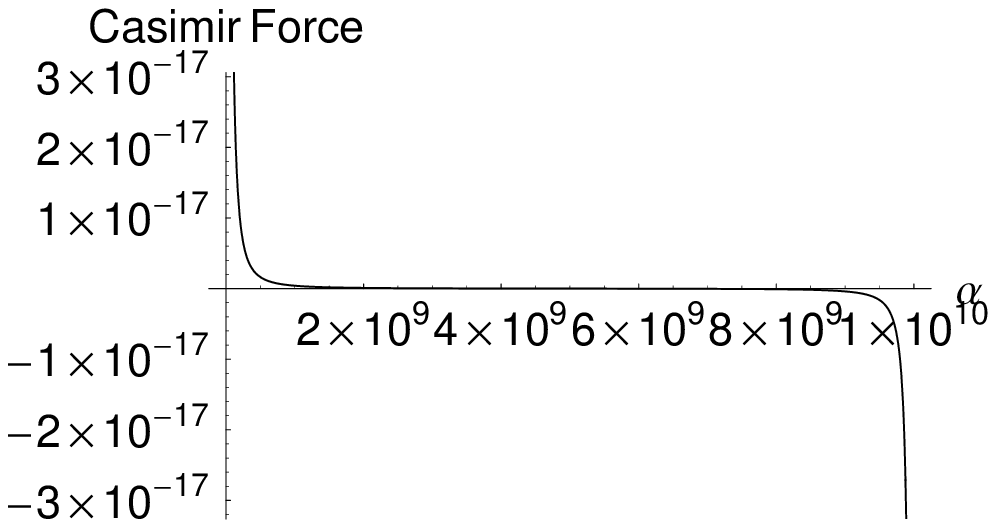}
\end{center}
\caption{The total Piston Casimir Force, for $R=10^{100}$,
$L=10^{11}$ and for a wide range of $\alpha$ (\textbf{Piston 2
Case})} \label{CasimirFT1}
\end{figure}

\section{Brief Discussion}

Let us discuss the results of the previous section. We start with
the piston 1 that appears in Figure \ref{MyPiston}. The Casimir
force that stems out of this configuration is described by Figures
\ref{CasimirForce1}, \ref{CasimirForce2} and
\ref{CasimirForceTotal}. As it can be seen, the total Casimir
force is negative for small $\alpha$. Also for large $\alpha$, for
values near $L$, the total Casimir energy is positive. In
conclusion the Casimir force is attractive when the piston is near
to the one end. In addition it is repulsive if the piston goes to
the other end. This kind of behavior is a known result for pistons
(see \cite{KirstenKaluzaKleinPiston}). Remember this case
corresponds to a piston that is ''free to move'' along one of the
non curved dimensions.

\noindent In the case of Piston 2 of Figure \ref{MyPiston2} the
results are different. Now the piston is ''free to move'' along
the slightly curved dimension of total length $L$. This case is
best described by Figures \ref{CasimirF1}, \ref{CasimirF2} and
\ref{CasimirFT1}. As we can see, the total Casimir force is
positive for small $\alpha$ values and negative for large $\alpha$
values (but still smaller than $L$). This means that the Casimir
force is repulsive to the one end and attractive to the other one.
Note that this behavior is similar to the Piston 1 case with the
difference that the force is attractive (repulsive) to different
places. However the qualitative behavior that is described by
repulsion to one end and attraction to the other, still holds. In
the next section we shall verify this using a semi-analytic
approximation.

\section{A Semi-analytic Approach for the Piston Casimir Energy and Casimir Force}

In this section we shall consider as in previous sections a three
dimensional piston $R^3$ and for simplicity the piston of Figure
\ref{MyPiston2}. Our study will be focused on the semi-analytic
calculation of the Casimir energy and Casimir force.

\noindent The piston configuration consists of two chambers with
lengths $\alpha$ and $L-\alpha$, and with Dirichlet boundary
conditions on the boundaries and on the moving piston. The energy
eigenfrequency for this setup is:
\begin{equation}\label{pistoneigenreg}
\omega ^2=k^2_1+k_2^2+(\frac{n\pi}{\alpha})^2-y^2,
\end{equation}
for the $\alpha$ chamber, and
\begin{equation}\label{pistoneigernr}
\omega ^2=k^2_1+k_2^2+(\frac{n\pi}{L-\alpha})^2-y^2,
\end{equation}
for the $L-\alpha$ chamber. In the above two relations the
parameter $y$ stands for a positive number with physical
significance analogous to the ones we described in the previous
sections. The Casimir energy with no regularization for this
system is given by:
\begin{equation}\label{totalcasimirenergy1}
E_{Piston}=E_{P}(L-\alpha)+E_{P}(\alpha),
\end{equation}
where $E_{P}(\alpha)$ is (upon integrating the infinite
dimensions):
\begin{equation}\label{unregcasimir}
E_{P}(\alpha)=\frac{3}{2\pi}\sum_{n=1}^{\infty}\Big{[}\big{(}\frac{n\pi}{\alpha}\big{)}^2-y^2\Big{]}^{3/2}.
\end{equation}
As we said, the above sum contains a singularity and needs
regularization. In order to see how the singularity ''behaves'' we
use the binomial expansion (or a Taylor expansion for small $y$,
which is exactly the same as can be checked):
\begin{equation}\label{94}
(a^{2}-b^{2})^{s}=\sum_{l=0}^{\sigma
}\frac{s!}{(s-l)!l!}(a^{2})^{l}(-b^{2})^{s-l},
\end{equation}
and rearranging the sum as:
\begin{equation}\label{unregcasimir1}
E_{P}(\alpha)=\frac{3\pi^2}{2\alpha^3}\sum_{n=1}^{\infty}(n^2-m^2)^{3/2},
\end{equation}
and $m=\frac{y\alpha}{\pi}$ we obtain:
\begin{align}\label{taylorexpansions}
E_{P}(\alpha)=&\sum_{n=1}^{\infty}\frac{3}{2\pi}\Big{(}\frac{n^3}{\alpha^3}-\frac{3y^2n}{2\alpha
\pi^2} +\frac{3\alpha y^4}{8n\pi
^4}+\frac{\alpha^3y^6}{16n^3\pi^6}\\ \notag
&+\frac{3\alpha^5y^6}{128n^5\pi^8}+\frac{3\alpha^7y^{10}}{256n^7\pi^{10}}+\frac{7\alpha^9y^12}{1024n^9\pi^12}...\Big{)}.
\end{align}
Using the zeta regularization method \cite{Elizalde,eli} the above
relation (\ref{taylorexpansions}) becomes,
\begin{align}\label{taylorexpansions1}
E_{P}(\alpha)=&\frac{3}{2\pi}\Big{(}\frac{\zeta
(-3)}{\alpha^3}-\frac{3y^2\zeta (-1)}{2\alpha \pi^2}+\frac{3\alpha
y^4\zeta (1)}{8\pi ^4}+\frac{\alpha^3y^6\zeta (3)}{16\pi^6}\\
\notag &+\frac{3\alpha^5y^6\zeta
(5)}{128\pi^8}+\frac{3\alpha^7y^{10}\zeta
(7)}{256\pi^{10}}+\frac{7\alpha^9y^{12}\zeta
(9)}{1024\pi^{12}}...\Big{)}.
\end{align}
Notice that the singularity is contained in $\zeta (1)$ and is a
pole of first order. Now the other chamber of the piston
contributes to the Casimir energy as:
\begin{align}\label{taylorexpansions1la}
E_{P}(L-\alpha)=&\frac{3}{2\pi}\Big{(}\frac{\zeta
(-3)}{(L-\alpha)^3}-\frac{3y^2\zeta (-1)}{2(L-\alpha)
\pi^2}+\frac{3(L-\alpha) y^4\zeta (1)}{8\pi
^4}+\frac{(L-\alpha)^3y^6\zeta
(3)}{16\pi^6}\\
\notag &+\frac{3(L-\alpha)^5y^6\zeta
(5)}{128\pi^8}+\frac{3(L-\alpha)^7y^{10}\zeta
(7)}{256\pi^{10}}+\frac{7(L-\alpha)^9y^{12}\zeta
(9)}{1024\pi^{12}}...\Big{)}.
\end{align}
Before proceeding we must mention that we could reach the same
results as above by Taylor expanding the initial sum of relation
(\ref{unregcasimir}) for $y\rightarrow 0$.

\noindent Let us now discuss the above. It can be easily seen that
the total Casimir force,
\begin{equation}\label{tcf}
F_c=-\frac{\partial E_{P}(L-\alpha)}{\partial
\alpha}-\frac{\partial E_{P}(\alpha)}{\partial \alpha},
\end{equation}
is free or singularity. The reason is obvious and it is because
the pole containing term is linear to the length of the piston
chamber. Due to this linearity, the derivative of the energy
cancels this dependence and the two singularities cancel each
other. Thus we see how in our case also, a slightly curved piston
configuration results to a singularity free Casimir force.

\noindent Finally, let us discuss something different. One
quantity that is also free of singularity is,
\begin{equation}\label{newconce}
N(\alpha)=
\frac{E_P(\alpha)}{\alpha}-\frac{E_P(L-\alpha)}{L-\alpha}.
\end{equation}
It is not accidental that this quantity has dimensions Energy per
length which is actually force. The finiteness of the piston
Casimir force and the force related quantity $N(\alpha)$ show once
more, as is well known \cite{calvacanti,KirstenKaluzaKleinPiston},
that the piston configuration has many attractive field theoretic
features, even in the presence of very small curvature in one of
the dimensions. The same arguments hold regarding the quantity
(\ref{newconce}), also when we consider a massive scalar field in
an usual $R^3$ piston configuration (with no curved dimensions).

\subsection{Semi-analytic Analysis for the Piston Casimir Force}

\begin{figure}[h]
\begin{center}
\includegraphics[scale=.6]{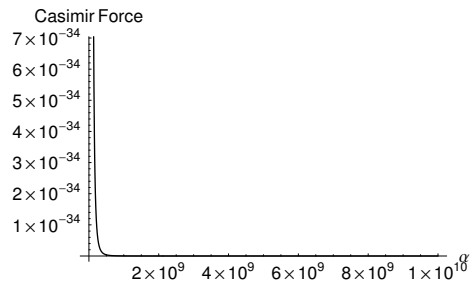}
\end{center}
\caption{The Casimir Force for the chamber $\alpha$, for
$R=10^{20}$, $L=10^{11}$ (\textbf{Piston 2 Case})}
\label{AnalyticForceA}
\end{figure}

\begin{figure}[h]
\begin{center}
\includegraphics[scale=.6]{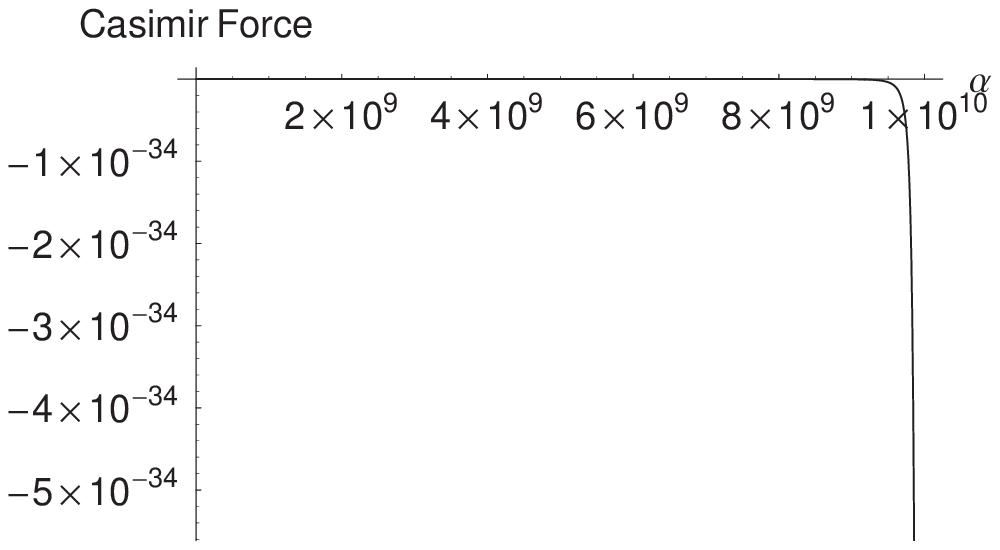}
\end{center}
\caption{The total Piston Casimir Force, for $R=10^{20}$,
$L=10^{11}$ (\textbf{Piston 2 Case})} \label{AnalyticLA}
\end{figure}

\begin{figure}[h]
\begin{center}
\includegraphics[scale=.6]{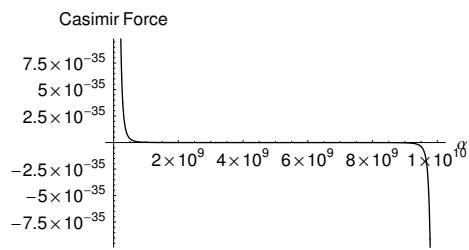}
\end{center}
\caption{The total Piston Casimir Force, for $R=10^{20}$,
$L=10^{11}$ and for a wide range of $\alpha$ (\textbf{Piston 2
Case})} \label{AnalyticTotalForce}
\end{figure}

According to the above, the Casimir energy can be approximated by
relations (\ref{taylorexpansions1}) and
(\ref{taylorexpansions1la}). Thus when the radius of the slightly
curved dimension is very large, the Casimir force can be
approximated by,
\begin{equation}\label{forceanalytic}
F_c(\alpha)=-\frac{9\pi \zeta(-3)}{\alpha
^4}+\frac{9y^2\zeta{(-1)}}{2\alpha ^2\pi}+\frac{9\alpha
^2y^6\zeta{(3)}}{16\pi^5},
\end{equation}
for the chamber $\alpha$, and
\begin{equation}\label{forceanalytic}
F_c(L-\alpha)=-\frac{9\pi \zeta(-3)}{(L-\alpha)
^4}+\frac{9y^2\zeta{(-1)}}{2\alpha ^2\pi}+\frac{9(L-\alpha)
^2y^6\zeta{(3)}}{16\pi^5},
\end{equation}
for the $L-\alpha$ chamber. When $R$ is much larger than $\alpha$,
(with $y=1/R$) then the approximation we used above is adequate to
describe the Casimir force. In Figures \ref{AnalyticForceA},
\ref{AnalyticLA} and \ref{AnalyticTotalForce}, we plot the Casimir
force for the chamber $\alpha$, $L-\alpha$ and the total force,
for $R=10^{20}$ and $L=10^{11}$
As it can be seen from the Figures, the behavior of the Casimir
force is the same as the one we found in section 1. Thus for the
Piston 2 configuration, the Casimir force is positive for the
$\alpha$ chamber and negative for the $L-a$ chamber (when $\alpha$
approaches the size of $L$). Also the total Casimir force is
positive for small $\alpha$ and negative for large $\alpha$.

\noindent A similar analysis can be carried for the Piston 1
configuration and similar results hold.

\begin{figure}[h]
\begin{center}
\includegraphics[scale=.6]{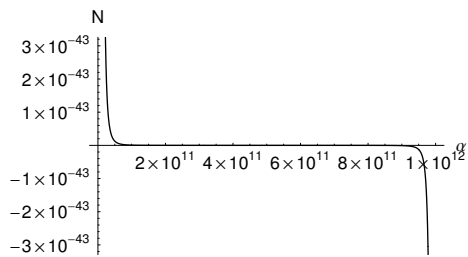}
\end{center}
\caption{The quantity $N(\alpha)$, for $R=10^{100}$, $L=10^{12}$
and for a wide range of $\alpha$ (\textbf{Piston 2 Case})}
\label{newconce12}
\end{figure}

\noindent Finally we plot the quantity $N(\alpha)$ of relation
(\ref{newconce}). In Figure \ref{newconce12} we can see that the
behavior of $N(\alpha)$ resembles that of the total Casimir force
we studied previously, for the case of Piston 2, and for the
$\alpha$ chamber.

\noindent Before closing this section let us briefly comment on
the regularization dependence of the force corresponding to the
pistons we used. We shall make use of a cutoff parameter $\lambda$
in the Casimir energy, in order to see explicitly the cancellation
of the surface terms we mentioned in the introduction
\cite{Bordagreview,Teo,teo1,teo2,edery1,ar1,ar2,ar3}. We work
again for the piston 2 configuration. The Casimir energy for the
$a$ chamber is \cite{Teo,teo1,teo2},
\begin{equation}\label{lreg}
E^P_{Cas}(\lambda,a,y)=\frac{1}{2}\sum_{n=1}^{\infty}\frac{3}{2\pi}\Big{[}\big{(}\frac{n\pi}{\alpha}\big{)}^2-y^2\Big{]}^{3/2}e^{-\lambda
\Big{[}\big{(}\frac{n\pi}{\alpha}\big{)}^2-y^2\Big{]}^{3/2}}
\end{equation}
The finite Casimir energy is obtained by taking the limit $\lambda
\rightarrow 0$ in the regular term of the Casimir energy (we will
see it shortly). After using the inverse Mellin transform
\cite{Teo,teo1,teo2} of the exponential and observing that,
\begin{equation}\label{pppp}
E^P_{Cas}(\lambda,a,y)=-\frac{3}{2}\frac{\partial }{\partial
\lambda}\sum_{n=1}^{\infty}e^{-\lambda
\Big{[}\big{(}\frac{n\pi}{\alpha}\big{)}^2-y^2\Big{]}^{3/2}}
\end{equation}
we obtain (following the technique of reference
\cite{Teo,teo1,teo2}),
\begin{equation}\label{regenergrgy}
E^P_{Cas}(\lambda,a,y)=a\frac{3}{2\pi
\lambda^2}+\frac{3{\,}y^2{\,}a}{4\pi}\log \lambda
+E_{\mathrm{regular}}(a)
\end{equation}
for the $a$ chamber, while for the $L-a$ chamber we obtain
accordingly,
\begin{equation}\label{regenergrgy221}
E^P_{Cas}(\lambda,L-a,y)=(L-a)\frac{3}{2\pi
\lambda^2}+\frac{3{\,}y^2{\,}(L-a)}{4\pi}\log \lambda
+E_{\mathrm{regular}}(L-a)
\end{equation}
As it can be easily seen, the Casimir force is free of
singularities because these cancel when we add the two
contributions. Indeed,
\begin{equation}\label{regenergrgy23}
F^P_{Cas}(\lambda,a,y)=\frac{3}{2\pi
\lambda^2}+\frac{3{\,}y^2{\,}}{4\pi}\log \lambda
+F_{\mathrm{regular}}(a)
\end{equation}
and
\begin{equation}\label{regenergrgy221}
F^P_{Cas}(\lambda,L-a,y)=-\frac{3}{2\pi
\lambda^2}-\frac{3{\,}y^2}{4\pi}\log \lambda
+F_{\mathrm{regular}}(L-a)
\end{equation}
Thus the total Casimir force is regular,
\begin{equation}\label{tottttt}
F_{\mathrm{total}}=F_{\mathrm{regular}}(a)+F_{\mathrm{regular}}(L-a)
\end{equation}

\section*{Conclusions}

In this paper we studied the Casimir force for two piston
configurations. We used Dirichlet boundary conditions and the
pistons had a slightly curved dimension. We found the
eigenfunctions (in first order approximation with respect to $R$)
and the $R$-dependent eigenvalues, for two piston configurations.
These configurations appear in Figures \ref{MyPiston} and
\ref{MyPiston2}. For the case of Piston 1, we found that the
Casimir force is attractive when the piston is near to the one end
and particularly in the end which is near to $\alpha \rightarrow
0$. In addition it is repulsive if the piston goes to the other
end. Remember this case corresponds to a piston that is ''free to
move'' along one of the non-curved dimensions.

\noindent In the case of Piston 2 (Figure \ref{MyPiston2}) the
results are different. Now the piston is free to move along the
slightly curved dimension of total length $L$. The total Casimir
force is positive for small $\alpha$ values and negative for large
$\alpha$ values (but still smaller than $L$). This means that the
Casimir force is repulsive to the one end and attractive to the
other one. Note that this behavior is similar to the Piston 1 case
with the difference that the force is attractive (repulsive) to
different places.

\noindent This said behavior, that is, the Casimir force on the
piston being attractive in the one end and repulsive to the other,
is a feature of Casimir pistons, see
\cite{KirstenKaluzaKleinPiston}. We verified this behavior
following a semi-analytic method. Also we found that the quantity,
\begin{equation}\label{newconce1}
N(\alpha)=
\frac{E_P(\alpha)}{\alpha}-\frac{E_P(L-\alpha)}{L-\alpha},
\end{equation}
which has dimensions of force, has the same behavior as the piston
force for the same chamber $\alpha$. Notice that $E_P(\alpha)$
appears first in $N(\alpha )$ and $E_P(L-\alpha)$ is subtracted
from it.

\noindent In conclusion we saw how a slightly curved dimension
alters the Casimir force for a Casimir piston with Dirichlet
boundary conditions. It would be interesting to add the
contribution of an extra dimensional space. Particularly a three
dimensional compact manifold, since in most cases the predictions
of ADD models for large extra dimensions corrections of the Newton
law, rule out manifolds \cite{kribs} with dimensions less than 3
(of course with $TeV$ compactification scale). Indeed in table 1
this is seen clearly.
\begin{center}
\begin{tabular}{|c|c|}
  \hline
  number of extra dimensions & R (m) \\
  \hline
  n=1 & $\sim 10^{12}$ \\
  \hline
  n=2 & $\sim 10^{-3}$ \\
  \hline
  n=3 & $\sim 10^{-8}$ \\
\hline
\end{tabular}
\end{center}
In the setup we used in this paper, the incorporation of a Ricci
flat manifold or a positive curved manifold could be done using
standard techniques
\cite{KirstenKaluzaKleinPiston,Teo,teo1,teo2,Cheng}. One
interesting case would involve hyperbolic manifolds and especially
with non Poisson spectrum of their eigenvalues. Also a single
$R^3$ piston configuration with one curved dimension could not
support Neumann boundary conditions for the scalar field. However
such a configuration with an extra dimensional structure could
hold if the extra dimensional space has non zero index or if it
has an orbifold structure. We shall report on these issues soon.

\section*{Acknowledgements}

The author is indebted to Prof. K. Kokkotas and his Gravity Group
for warm hospitality in Physics Dept. Tuebingen where part of this
work was done. Also the author would like to thank N. Karagianakis
and M. Koultouki for helping him with the figures.

\end{document}